\documentclass[aps,prb,twocolumn,amsmath,amssymb,nofootinbib,superscriptaddress,floatfix]{revtex4}

\usepackage{amsmath}
\usepackage{amssymb}
\usepackage{amsthm}
\usepackage[dvips]{color} %Christopher
\usepackage{graphicx}% Include figure files
\usepackage{dcolumn} % Align table columns on decimal point
\usepackage{bm} % bold math
\usepackage[hypertex]{hyperref}
\usepackage{longtable}
\usepackage{ulem}   % to strike things out
\begin{document}

\title{
Topological qubits in graphenelike systems
      }

\author{Luiz Santos} 
\affiliation{
Department of Physics, 
Harvard University, 
17 Oxford Street, 
Cambridge, Massachusetts 02138,
USA
            } 
\author{Shinsei Ryu} 
\affiliation{Department of Physics, University of California,
Berkeley, California 94720, USA}

\author{Claudio Chamon} 
\affiliation{
Physics Department, 
Boston University, 
Boston, Massachusetts 02215, USA
            } 
\author{Christopher Mudry} 
\affiliation{
Condensed Matter Theory Group, 
Paul Scherrer Institute, CH-5232 Villigen PSI,
Switzerland
            } 

\date{\today}

\begin{abstract}
The fermion-doubling problem can be an obstacle to getting half a qubit
in two-dimensional fermionic tight-binding models in the form of 
Majorana zero modes bound to the core of superconducting vortices.
We argue that the number of such Majorana zero modes is determined by a 
$\mathbb{Z}^{\ }_{2}\times\mathbb{Z}^{\ }_{2}$ 
topological charge for a family of two-dimensional
fermionic tight-binding models
ranging from noncentrosymmetric materials to graphene.
This charge depends on the dimension of the representation
(i.e., the number of species of Dirac fermions -- where the
doubling problem enters) and the parity of the Chern number induced by
breaking time-reversal symmetry. We show that in graphene there are as
many as ten order parameters that can be used in groups of four 
to change the topological number from even to odd. 
\end{abstract}
\maketitle

\section{
Introduction
        }

A major hurdle in the realization of quantum computers is the problem
of decoherence. 
Qubits generically do not last long in the presence of the environment. 
Overcoming decoherence is possible if the qubit is stored 
nonlocally using a many-body state, in such a way that the reservoir, 
which only couples locally to the system, is unable to damage the 
quantum information.~\cite{Kitaev03}
An implementation of this scheme
can be achieved if the many-body ground state supports excitations 
obeying non-Abelian braiding statistics. Non-Abelian braiding statistics
that departs from the Bose-Einstein or Fermi-Dirac statistics
can only be realized in effectively two-dimensional (2D) systems,
such as the $\nu=5/2$ fractional quantum Hall state
on the one hand,~\cite{Moore91} or in chiral 
$p^{\ }_{x}\pm i p^{\ }_{y}$ 
2D superconductors where a half vortex binds a zero-energy 
midgap state on the other hand.~\cite{Read00,Ivanov01} 
In the latter example, due to particle-hole symmetry (PHS), 
this zero mode is a Majorana fermion. A two-state system
-- a qubit -- can be assembled from one complex fermion made up of a
pair of such Majorana fermions sitting at far away vortices. The
splitting of the energies from exactly zero is exponentially small in
the separation between the vortices and thus the time to degrade a
quantum superposition is exponentially large in the distance between
the vortices.

Majorana fermions bound to the core of vortices in a superconductor
were discovered by Jackiw and Rossi.~\cite{Jackiw81} There
is a single zero-energy midgap state when the vorticity is $n=\pm 1$
(there are generically $|n|$ zero modes). However, these results are
obtained for the case of the minimal representation of
Dirac fermions whose support in space is
the 2D continuum.
The fermion-doubling problem, discovered in the context of lattice-gauge theory,~\cite{Nielson81} 
prevents importing these results of Jackiw and Rossi
to condensed-matter systems, which are models defined on lattices. 
In graphene, for instance, one does have the Majorana fermions 
from which it is possible to assemble the qubits, as shown by Ghaemi and
Wilczek.~\cite{Ghaemi07} However, there are four of them in each
vortex because there are two Dirac cones for each spin polarization
in graphene. Even numbers of Majorana fermions are not stable, as local
perturbations can move them away from zero energy.

One can get much insight into the problem of how many Majorana fermions 
can be realized in effectively 2D tight-binding models
if one looks into ideas for addressing the fermion-doubling problem
in lattice-gauge theories. There is the original proposal due to Wilson,
which is achieved by adding perturbations (Wilson masses)
to a lattice
Hamiltonian that opens gaps at undesirable duplicate Dirac
points.~\cite{Wilson77} However, from the point of view of a
lattice-gauge regularization of quantum chromodynamics,
this option has the undesirable property of breaking the chiral symmetry.
Alternatively, the idea that the fermion doubling 
can be overcome by considering an $n$-dimensional system as a boundary 
of a $(n+1)$-dimensional one  was put forth by Callan and Harvey in 
Ref.~\onlinecite{Callan85} 
(see also Refs.~\onlinecite{Fradkin86} and \onlinecite{Kaplan92}). 
In fact, it is the Callan-Harvey effect that is at work 
in the remarkable results obtained by Fu and Kane:~\cite{Fu08}
(i)
surface states in 3D topological insulators with time-reversal symmetry 
(TRS) realize an odd number of Dirac fermions in the minimal representation
of the Clifford algebra. 
(ii) They can be used to achieve an odd number of
Majorana fermions bound to vortex cores induced by the
proximity to a type-II superconductor.

We argue in this paper that the Wilson prescription is a route to
attain an odd number of Majorana fermions in effectively 2D
condensed-matter systems. The chief reason is that one is not constrained 
to impose chiral symmetry as in lattice-gauge theory. 
However, not any Wilson mass can be used for this
purpose, only those that break TRS. This
approach naturally leads to a
$\mathbb{Z}^{\ }_{2}\times\mathbb{Z}^{\ }_{2}$ 
topological charge that discerns
whether the system has an even or odd number of Majorana fermions
attached to a superconducting (SC) vortex. In essence, the parity of the
number of zero modes is determined by the number of Dirac points which
have not been knocked out by changing the Chern number via the
TRS-breaking Wilson mass. For systems where the number of species is
odd, like in the case of surface states of TRS topological insulators,
odd numbers of zero modes occur without breaking TRS (odd$\times$even
case). In the case of graphene, which is one focus of this paper and where
there is an even number of Dirac cones, TRS must be broken so as
to obtain an odd Chern number and, in turn, an odd number of
Majorana fermions (even$\times$odd case). In all other cases, including
graphene when TRS is unbroken (even$\times$even case) and surface
states of topological insulators with large enough magnetic field
(odd$\times$odd case), there are even numbers of Majorana fermions. 
Notice that, according to this
$\mathbb{Z}^{\ }_{2}\times\mathbb{Z}^{\ }_{2}$ 
classification, systems defined on 2D
lattices must have both SC pairing correlations and a
nonzero Chern number that accounts for 
a \textit{thermal} Hall effect in order to have
non-Abelian quasiparticles. This is ``the poor cousin,'' i.e., the
mean-field version, of the $\nu=5/2$ quantum Hall state.

%%%%%% BEGIN FIGURE
\begin{figure}[tb]
\begin{center}
\includegraphics[angle=0,scale=0.65]{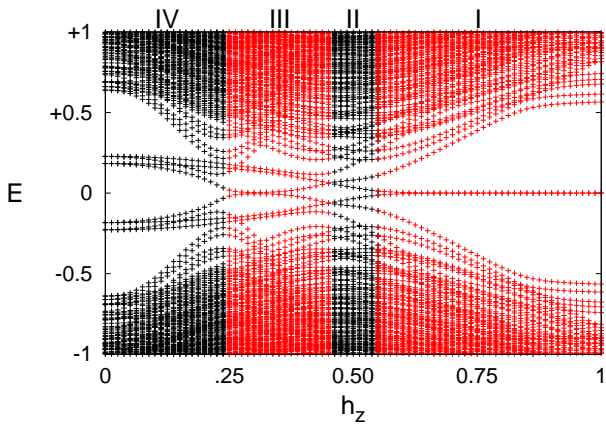}
\caption{
(Color online) 
Energy levels for the lattice Hamiltonian%
~(\ref{eq: def Hamiltonian H}) 
as a function of 
$h^{\ }_{3}\equiv h^{\ }_{z}$ 
for 
$h^{\ }_{1}\equiv h^{\ }_{\hat{\boldsymbol{x}}}=0.55 h^{\ }_{3}$,
$h^{\ }_{2}\equiv h^{\ }_{\hat{\boldsymbol{y}}}=0.45 h^{\ }_{3}$,
and
$\alpha=\Delta(r=\infty)=1$. 
Although finite-size effects prevent the closing of the bulk gap, 
the vanishing curvature of the gap in the bulk is a footprint of 
each thermodynamic transition as a function of $h^{\ }_{z}$.
In region IV, there are 8 (=4$\times2$) midgap states, 
four bound to the vortex and four bound to the antivortex.
Their degeneracy is lifted by the combined effects
of intravortex-level or intervortex-level repulsion.
There are 6 (=3$\times$2), 4 (=2$\times$2), and 2 (=1$\times$2) 
midgap states in regions III, II, and I, respectively.
In the limit in which the vortex and antivortex separation 
goes to infinity, the intervortex-level repulsion is 
exponentially suppressed and
midgap states can be converted to Majorana fermions.
Thus, an odd number of Majorana fermions attached to an isolated vortex
are found in regions III and I. 
        }
\label{fig:numerics}
\end{center}
\end{figure}
%%%%%% END FIGURE

\section{
Tuning the number of Majorana fermions
        }

To illustrate how it is possible to use a Wilson mass prescription
to change at will the number of Majorana fermions in a 2D
tight-binding model, we begin with the pure Rashba kinetic energy
\begin{equation}
H^{\ }_{\alpha}:=
\sum_{\boldsymbol{r}}
\sum_{\hat{\boldsymbol{n}}=\hat{\boldsymbol{x}},\hat{\boldsymbol{y}}}
\left(
i\alpha\,
c^{\dag}_{\boldsymbol{r}+\hat{\boldsymbol{n}}}\;
\sigma^{\ }_{\hat{\boldsymbol{n}}}\;
c^{\   }_{\boldsymbol{r}\vphantom{+\hat{\boldsymbol{n}}}}
+
\mathrm{H.c.}
\right)
\label{eq: def H alpha}
\end{equation}
that we define on a square lattice with sites
denoted by $\boldsymbol{r}=m\hat{\boldsymbol{x}}+n\hat{\boldsymbol{y}}$,
where $m$ and $n$ are integers. Here, 
$c^{\dag}_{\boldsymbol{r}}=(c^{\dag}_{\boldsymbol{r},\mathrm{s}})$
is a doublet that creates on site $\boldsymbol{r}$ an electron with
the spin projection $\mathrm{s}=\uparrow,\downarrow$ along the 
quantization axis, 
$\sigma^{\ }_{\hat{\boldsymbol{x}}}\equiv
\sigma^{\ }_{x}\equiv\sigma^{\ }_{1}$
and
$\sigma^{\ }_{\hat{\boldsymbol{y}}}\equiv
\sigma^{\ }_{y}\equiv\sigma^{\ }_{2}$
are the first two Pauli matrices while the third
Pauli matrix
$\sigma^{\ }_{3}$ defines
the quantization axis in spin space, 
and the real-valued number $\alpha$ sets
the energy scale for the Rashba hopping. 
At half filling, i.e., at vanishing chemical potential,
the Fermi surface collapses to the four nonequivalent Fermi points 
\begin{equation}
%\boldsymbol{k}^{\ }_{\mathrm{F}}=
\boldsymbol{p}^{\ }_{\mathrm{F}}=
(0,0),
(0,\pi),
(\pi,0),
(\pi,\pi).
\label{eq: def Fermi points}
\end{equation}
These Fermi points are TRS in that they change by
a reciprocal wave vector under the inversion 
%$\boldsymbol{k}\to-\boldsymbol{k}$.
$\boldsymbol{p}\to-\boldsymbol{p}$.
Linearization of the energy spectrum of 
$H^{\ }_{\alpha}$ in the vicinity of these four Fermi points yields
an $8\times8$ massless Dirac Hamiltonian, i.e., 
a reducible representation of the Clifford algebra
four times larger than the minimal one in 2D continuum space.
This is a manifestation of the fermion doubling.
Hamiltonian~(\ref{eq: def H alpha}) preserves TRS but breaks
completely SU(2) spin-rotation symmetry (SRS).
% and inversion
%symmetry $\boldsymbol{k}\to-\boldsymbol{k}$.

We now introduce the same spectral gap at all the Fermi points,
Eq.~(\ref{eq: def Fermi points}).
We achieve this with the help of a singlet SC order
parameter parametrized by a complex-valued $\Delta$,
\begin{equation}
H^{\ }_{\Delta}:=
\sum_{\boldsymbol{r}}
\left[
\Delta\,
c^{\dag}_{\boldsymbol{r}\vphantom{\hat{\boldsymbol{y}}}}\,
\left(
i\sigma^{\ }_{\hat{\boldsymbol{y}}}
\right)\,
c^{\dag}_{\boldsymbol{r}\vphantom{\hat{\boldsymbol{y}}}}                    
+
\mathrm{H.c.}
\right],
\label{eq: def H Delta}
\end{equation}
that we add to Hamiltonian~(\ref{eq: def H alpha}),
\begin{equation}
H^{\ }_{\alpha}\to H^{\ }_{\alpha}+H^{\ }_{\Delta}.
\end{equation}
This 2D tight-binding Hamiltonian is the relative to
a noncentrosymmetric superconductor in the Rashba-Dirac limit
and with singlet SC pairing studied in Ref.~\onlinecite{Santos09}.
A TRS-breaking vortex
\begin{equation}
\Delta\to\Delta(r)\exp({i}\theta)
\label{eq: def vortex}
\end{equation}
with the profile  $\Delta(r)$, where
$r$ and $\theta$ are polar coordinates,
binds four Majorana fermions at the Fermi energy
via the Jackiw-Rossi solutions.~\cite{Santos09}

Finally, we define the three independent
TRS-breaking Wilson masses 
\begin{equation}
H^{\ }_{\boldsymbol{h}}:=
\sum_{\boldsymbol{r}}
\left(
\sum_{\hat{\boldsymbol{n}}=\hat{\boldsymbol{x}},\hat{\boldsymbol{y}}}
h^{\ }_{\hat{\boldsymbol{n}}\vphantom{+\hat{\boldsymbol{n}}}}\,
c^{\dag}_{\boldsymbol{r}+\hat{\boldsymbol{n}}}\,
\sigma^{\vphantom{\dag}}_{3\vphantom{+\hat{\boldsymbol{n}}}}
c^{\vphantom{\dag}}_{\boldsymbol{r}\vphantom{+\hat{\boldsymbol{n}}}}
+
\mathrm{H.c.}
+
2
h^{\vphantom{\dag}}_{3\vphantom{+\hat{\boldsymbol{n}}}}\,
c^{\dag}_{\boldsymbol{r}\vphantom{+\hat{\boldsymbol{n}}}}\,
\sigma^{\vphantom{\dag}}_{3\vphantom{+\hat{\boldsymbol{n}}}}
c^{\vphantom{\dag}}_{\boldsymbol{r}\vphantom{+\hat{\boldsymbol{n}}}}
\right)
\label{eq: def H vec h}
\end{equation}
parametrized by the 
triplet of energy scales, 
\begin{equation}
\boldsymbol{h}=
(h^{\ }_{1},h^{\ }_{2},h^{\ }_{3})\equiv
(h^{\ }_{\hat{\boldsymbol{x}}},h^{\ }_{\hat{\boldsymbol{y}}},h^{\ }_{z})\equiv
(h^{\ }_{x},h^{\ }_{y},h^{\ }_{z}).
\end{equation}
Each Wilson mass breaks TRS and breaks SU(2) SRS down to U(1).

We are going to show numerically on the lattice
and analytically in the continuum limit that the Wilson masses
$\boldsymbol{h}$
can be used to change  the number of Majorana fermions bound to
the core of the vortex, Eq.~(\ref{eq: def vortex}), in 
\begin{equation}
H:=
H^{\ }_{\alpha}
+
H^{\ }_{\Delta}
+
H^{\ }_{\boldsymbol{h}}
\label{eq: def Hamiltonian H}
\end{equation}
one by one from four to zero.

Figure~\ref{fig:numerics} displays the energy spectrum of
Hamiltonian~(\ref{eq: def Hamiltonian H}) obtained from numerical
diagonalization on a square lattice made of 39$\times$39 sites.
Periodic boundary conditions are imposed in the presence of a vortex
and an anti-vortex with winding numbers $\pm1$, respectively, that are
as far apart as possible.  A continuum of energy eigenstates is
visible as are bulk-gap-closing transitions as a function of
$h^{\ }_{3}\equiv h^{\ }_{z}$. Midgap states are also visible
although they are not located at the Fermi energy
because of level repulsion for states attached to the same defect and
because of the finite separation between the two defects. Starting
from four midgap states per isolated vortex in the thermodynamic limit,
increasing $h^{\ }_{3}$ decreases their number by
one after each bulk-gap-closing transition.

The same results follow analytically after linearization of the
spectrum around the Fermi points, Eq.~(\ref{eq: def Fermi points}).
Indeed, linearization of Hamiltonian~(\ref{eq: def Hamiltonian H})
yields, in the Bogoliubov-de-Gennes single-particle representation, 
the 16$\times$16-dimensional block-diagonal Hermitian matrix, 
\begin{subequations}
\label{eq: continuum limit noncentrosymmetric}
\begin{equation}
\mathcal{H}:=
\mathrm{diag} 
\begin{pmatrix}
\mathcal{H}^{\ }_{1},
\mathcal{H}^{\ }_{2},
\mathcal{H}^{\ }_{3},
\mathcal{H}^{\ }_{4}
\end{pmatrix}
\label{eq: continuum limit noncentrosymmetric a}
\end{equation}
with the 4$\times$4 Hermitian blocks
\begin{equation}
\mathcal{H}^{\ }_{j} = 
\begin{pmatrix}
-\eta^{\ }_{j} 
& 
p
& 
\delta^{\ }_{j} 
& 
0 
\\
\overline{p}
& 
\eta^{\ }_{j} 
& 
0 
& 
\delta^{\ }_{j} 
\\
\overline{\delta^{\ }_{j}} 
& 
0 
& 
-\eta^{\ }_{j} 
& 
-p
\\
0 
& 
\overline{\delta^{\ }_{j}} 
& 
-\overline{p}
& 
\eta^{\ }_{j}
\end{pmatrix}
\label{eq: continuum limit noncentrosymmetric b}
\end{equation}
whereby the units $\hbar=v^{\ }_{\mathrm{F}}=1$ have been chosen, 
the complex notation $p\equiv p^{\ }_{1}+{i}p^{\ }_{2}$ 
is used for the momenta whereby
$\overline{x}$ denotes the complex conjugate of $x$,
and
\begin{eqnarray}
&&
\delta^{\ }_{1}=\delta^{\ }_{2}\equiv
\Delta,
\qquad
\delta^{\ }_{3}=\delta^{\ }_{4}\equiv
\overline{\Delta},
\\
&&
\eta^{\ }_{1}\equiv
\eta^{\ }_{(0,0)}=
2\left(h^{\ }_{3}+h^{\ }_{1}+h^{\ }_{2}\right),
\\
&&
\eta^{\ }_{2}\equiv
\eta^{\ }_{(0,\pi)}=
-2\left(h^{\ }_{3}+h^{\ }_{1}-h^{\ }_{2}\right),
\\
&&
\eta^{\ }_{3}\equiv
\eta^{\ }_{(\pi,\pi)}=
2\left(h^{\ }_{3}-h^{\ }_{1}-h^{\ }_{2}\right),
\\
&&
\eta^{\ }_{4}\equiv
\eta^{\ }_{(\pi,0)}=
-2\left(h^{\ }_{3}-h^{\ }_{1}+h^{\ }_{2}\right).
\end{eqnarray}
\end{subequations}
The mathematical form of any of the four 
matrices $\mathcal{H}^{\ }_{j}$ is the same as that studied
in Ref.~\onlinecite{Ryu09}, provided the SC order parameter
$\Delta$ and each $\eta^{\ }_{j}$ are identified, respectively, with
the Kekul\'e and Haldane masses in Ref.~\onlinecite{Ryu09}. 
Thus, we can immediately borrow and tailor some
of the results from Ref.~\onlinecite{Ryu09}
to the present case.

If all the $\eta^{\ }_{j}$'s
are zero and the SC order parameter has a single vortex
with unit winding number, there are four Majorana fermions bound to it. 
As the magnitudes of
the $|\eta^{\ }_{j}|$'s increase, there will be a phase transition
every time that $|\eta^{\ }_{j}|=|\Delta(r=\infty)|$. 
Any such transition is
characterized by a decrease in the number of 
Majorana fermions by one unit and
a corresponding change in the value of the Chern number by $\pm 1$,
depending on the sign of $\eta^{\ }_{j}$. Therefore, by changing the Chern
number of the system by $\pm 1$ each time, one can knock out the
Majorana fermions one by one.
 
Alternatively, one could start from the dominant Haldane masses limit 
defined by $|\eta^{\ }_{j}|\,>\, |\Delta(r=\infty)|$ with $j=1,2,3,4$. 
In this limit, the system sustains the \textit{thermal}
integer quantum Hall effect (IQHE) 
and supports four chiral Majorana fermions.~\cite{Haldane88,footnotethermal} 
One can then change $\boldsymbol{h}$ so as to cross
successive quantum phase transitions at which any one of the
$\eta^{\ }_{1},\dots,\eta^{\ }_{4}$ 
equals in magnitude the spectral gap controlled by $|\Delta(r=\infty)|$.  
As before, each time we cross a phase transition, the Chern number and
hence the number of Majorana fermions at a SC vortex core
changes. 

We note that the presence of a nearest-neighbor-hopping dispersion
\begin{equation}
\epsilon(\boldsymbol{p})=-2t(\cos{p^{\ }_{x}}+\cos{p^{\ }_{y}})
\end{equation}
with $t\ll\alpha$ is equivalent to adding a constant chemical potential
\begin{equation}
\mu^{\ }_{j}\equiv\epsilon(\boldsymbol{p}^{\ }_{j})
\end{equation}
for each one of the 
four Fermi momenta, Eq.~(\ref{eq: def Fermi points}). The effect of this
term is to shift the gap closing condition to 
\begin{equation}
|\eta^{\ }_{j}|=\sqrt{|\Delta|^2+\mu^{2}_{j}},
\end{equation}
i.e., our results can be generalized to systems with quadratic dispersions
and naturally explain the results found in Ref.~\onlinecite{Sau09}.

\begin{table*}
\caption{ 
\label{table-1}
The ten mass matrices with PHS that
anticommute with $\alpha^{\ }_{1}$ and $\alpha^{\ }_{2}$ and
commute with the singlet SC masses 
$M^{\ }_{\text{ReSSC}}$
and 
$M^{\ }_{\text{ImSSC}}$.  Each mass matrix can be
assigned an order parameter for the underlying microscopic model. 
The latin subindex of the order parameter's name corresponds to the
preferred quantization axis in SU(2) spin space. Each mass matrix
either preserves or breaks TRS, SRS, and sublattice symmetry (SLS). Each mass matrix
can be written as a tensor product 
$X^{\ }_{\mu_1\mu_2\mu_3\mu_4}\equiv 
\rho^{\ }_{\mu^{\ }_1} 
\otimes 
s^{\ }_{\mu^{\ }_2} 
\otimes 
\sigma^{\ }_{\mu^{\ }_3} 
\otimes 
\tau^{\ }_{\mu^{\ }_4}$,
where 
$\rho^{\ }_{\mu^{\ }_1}$, 
$s^{\ }_{\mu^{\ }_2}$, 
$\sigma^{\ }_{\mu^{\ }_3}$, 
and 
$\tau^{\ }_{\mu^{\ }_4}$ 
correspond to unit $2\times2$ 
and Pauli matrices that act on particle-hole, spin-$1/2$,
valley,  and sublattice indices, respectively.
}
\begin{ruledtabular}
\begin{tabular}{llllll}
Order parameter & 
TRS & 
SRS & 
SLS &  
Here&
$X^{\ }_{\mu^{\ }_{1}\mu^{\ }_{2}\mu^{\ }_{3}\mu^{\ }_{4}}$\\
&&&&&\\ 
{IQHE} & 
{False}& 
{True} & 
{False}&
$M^{\ }_{\text{IQHE}}$&
$X^{\ }_{3003}$\\
&&&&&\\ 
{ReVBS}$^{\ }_{x}$& 
{False}& 
{False}& 
{True} &
$M^{\ }_{\text{ReVBS}^{\ }_{x}}$&
$X^{\ }_{3110}$\\
{ReVBS}$^{\ }_{y}$& 
{False}& 
{False}& 
{True} &
$M^{\ }_{\text{ReVBS}^{\ }_{y}}$&
$X^{\ }_{0210}$\\
{ReVBS}$^{\ }_{z}$& 
{False}& 
{False}& 
{True} &
$M^{\ }_{\text{ReVBS}^{\ }_{z}}$&
$X^{\ }_{3310}$\\
&&&&&\\ 
{ImVBS}$^{\ }_{x}$& 
{False}& 
{False}& 
{True} &
$M^{\ }_{\text{ImVBS}^{\ }_{x}}$&
$X^{\ }_{0120}$\\
{ImVBS}$^{\ }_{y}$& 
{False}& 
{False}& 
{True} &
$M^{\ }_{\text{ImVBS}^{\ }_{y}}$&
$X^{\ }_{3220}$\\
{ImVBS}$^{\ }_{z}$& 
{False}& 
{False}& 
{True} &
$M^{\ }_{\text{ImVBS}^{\ }_{z}}$&
$X^{\ }_{0320}$\\
&&&&&\\ 
{N\'eel}$^{\ }_{x}$& 
{False}& 
{False}& 
{False}&
$M^{\ }_{\text{N\'eel}^{\ }_{x}}$&
$X^{\ }_{3133}$\\
{N\'eel}$^{\ }_{y}$& 
{False}& 
{False}& 
{False}&
$M^{\ }_{\text{N\'eel}^{\ }_{y}}$&
$X^{\ }_{0233}$\\
{N\'eel}$^{\ }_{z}$& 
{False}& 
{False}& 
{False}&
$M^{\ }_{\text{N\'eel}^{\ }_{z}}$&
$X^{\ }_{3333}$\\
\end{tabular}
\end{ruledtabular}
\end{table*}

\section{
Tuning the number of Majorana fermions in graphene
        }

We are now going to demonstrate that the very same control on the number of 
Majorana fermions achieved with Hamiltonian%
~(\ref{eq: continuum limit noncentrosymmetric}) 
is also possible in graphene. We recall that in graphene,
electrons with spin $\mathrm{s}=\uparrow,\downarrow$ hop on a honeycomb lattice that is made of two
triangular sublattices A and B. The conduction and valence
bands touch at the two non-equivalent points 
$\boldsymbol{K}^{\ }_{\pm}$ located at the opposite corners
in the hexagonal first Brillouin zone 
(see Ref.~\onlinecite{graphene-review} for a review). 
Finally, to account for the
possibility of a SC instability, Nambu doublets are
introduced with the index p and h to distinguish particles
from their charge conjugate (holes). Hence, after linearization
of the spectrum about the Fermi points
$\boldsymbol{K}^{\ }_{\pm}$,
this leads to a single-particle
kinetic energy represented by a 16$\times$16-dimensional matrix
\begin{equation}
\mathcal{H}^{\ }_{\mathrm{D}}:=
\alpha^{\ }_{1}p^{\ }_{1}
+
\alpha^{\ }_{2}p^{\ }_{2}.
\end{equation}
Here, $\alpha^{\ }_{1}$ and $\alpha^{\ }_{2}$ are two
16$\times$16-dimensional Dirac matrices.

\begin{figure}[tb]
\begin{center}
\includegraphics[scale=0.8]{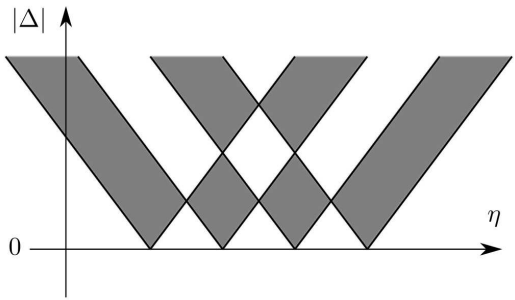}
\caption{
Schematic phase diagram of the competition between the 
singlet SC ($\Delta$), 
thermal IQH ($\eta$), 
magnetic bond ($m^{\ }_{1,2}$),
and  N\'eel ($m^{\ }_{3}$) orders near the Dirac point in graphene. 
Here, $m^{\ }_{1,2,3}$ are fixed while $|\eta|$ and $|\Delta|$ vary. 
When $m^{\ }_{1,2,3}=0$, there are three phases
(two large $|\eta|$ phases with $\eta>0$ and $\eta<0$
and one large $|\Delta|$ phase, according to Ref.~\onlinecite{Ryu09}).
If we choose $m^{\ }_{1,2,3}$ in such a way 
that
$m^{\ }_{1}+m^{\ }_{2}+m^{\ }_{3}$,
$m^{\ }_{1}+m^{\ }_{2}-m^{\ }_{3}$,
$-m^{\ }_{1}+m^{\ }_{2}-m^{\ }_{3}$,
and
$-m^{\ }_{1}-m^{\ }_{2}+m^{\ }_{3}$
are all different, there are 15 phases
as we change $\eta$ and $\Delta$. 
Shaded (nonshaded) regions represent 
a phase with the odd (even) Chern number 
(proportional to the thermal Hall conductivity
divided by temperature). 
In phases that are adiabatically connected to 
a nonsuperconducting state (the horizontal axis $|\Delta|=0$), 
one can switch off the pairing without closing the gap.
In these phases, the number of Majorana fermions is thus even. 
        }
\label{fig: phase_diagram}
\end{center}
\end{figure}

It was shown in Ref.~\onlinecite{Ryu09} that
there exists 36 distinct order parameters such that
any one, when added to $\mathcal{H}^{\ }_{\mathrm{D}}$,
opens a spectral gap. These order parameters are identified
by seeking all 16$\times$16 matrices from the Clifford algebra that 
anticommute with $\mathcal{H}^{\ }_{\mathrm{D}}$.
One complex valued order parameter is that for
singlet superconductivity. We shall denote the two
corresponding 16$\times$16 matrices from the Clifford algebra by
$M^{\ }_{\text{ReSSC}}$ 
and 
$M^{\ }_{\text{ImSSC}}$
and define the perturbation
\begin{equation}
\mathcal{H}^{\ }_{\Delta}:=
\Delta^{\ }_{1}
M^{\ }_{\text{ReSSC}}
+
\Delta^{\ }_{2}
M^{\ }_{\text{ImSSC}}
\end{equation}
that opens the spectral gap $2|\Delta|$
with the complex-valued
\begin{equation}
\Delta\equiv \Delta^{\ }_{1}+{i}\Delta^{\ }_{2}
\end{equation}
parametrized by the real-valued 
$\Delta^{\ }_{1}$ and $\Delta^{\ }_{2}$
when added to $\mathcal{H}^{\ }_{\mathrm{D}}$.
Next, we seek all 16$\times$16 matrices from the 
Clifford algebra that (i) anticommute with
$\mathcal{H}^{\ }_{\mathrm{D}}$
and (ii) commute with $\mathcal{H}^{\ }_{\Delta}$.
In this way, we find all ten TRS-breaking order parameters 
listed in Table~\ref{table-1}
that alone would open a gap in the Dirac spectrum
if not for their competition with the gap induced by
singlet superconductivity. 
Within this set of ten 
matrices one can form groups of at most four 
matrices that are mutually
commuting and therefore can be simultaneously diagonalized. Here, we
choose the four-tuplet 
$\{\text{ReVBS}^{\ }_{x},
   \text{ImVBS}^{\ }_{y}, 
   \text{N\'eel}^{\ }_{z}, 
   \text{IQHE} \}$ 
for concreteness but the results hereafter apply to any other such
four-tuplet of commuting mass matrices among the set of ten.
Our main result regarding graphene is the
fact that
\begin{subequations}
\begin{eqnarray}
\mathcal{H}&=& 
\boldsymbol{p} 
\cdot 
\boldsymbol{\alpha} 
+
\Delta^{\ }_{1}\, M^{\ }_{\text{ReSSC}}
+ 
\Delta^{\ }_{2}\, M^{\ }_{\text{ImSSC}}
\nonumber\\
&& {}
+ 
m^{\ }_{1}\, M^{\ }_{\text{ReVBS}^{\ }_{x}}
+ 
m^{\ }_{2}\, M^{\ }_{\text{ImVBS}^{\ }_{y}}
\nonumber\\
&& {}
+ m^{\ }_3\, M^{\ }_{\text{N\'eel}^{\ }_{z}} 
+ 
\eta\, M^{\ }_\text{IQHE}
\end{eqnarray}
is unitarily similar to 
Eqs.~(\ref{eq: continuum limit noncentrosymmetric a})
and
(\ref{eq: continuum limit noncentrosymmetric b})
with
\begin{eqnarray}
&&
\delta_{1,2,3,4}\equiv\Delta,
%\qquad
%\overline{\Delta}\equiv 
%\Delta_1 
%- 
%{i}\Delta_2,
\\
&&
\eta_1\equiv 
-m^{\ }_1+m^{\ }_2+m^{\ }_3+\eta,
\\
&&
\eta_2\equiv\, 
m^{\ }_1-m^{\ }_2+m^{\ }_3+\eta,
\\
&&
\eta_3\equiv\, 
m^{\ }_1+m^{\ }_2-m^{\ }_3+\eta,
\\
&&
\eta_4\equiv 
-m^{\ }_1-m^{\ }_2-m^{\ }_3+\eta.
\end{eqnarray}
\end{subequations}
The phase diagram in Fig.~\ref{fig: phase_diagram}
follows.

There is a total of four SC pair potentials that open a uniform gap at $\boldsymbol{K}^{\ }_{\pm}$ in graphene.%
~\cite{Ryu09}
One is the singlet SC pair potential, which we have discussed so far 
and the remaining three are all triplet SC pair potentials.
For each such triplet SC mass, as in the singlet SC,
there are four competing orders that commute pairwise
and can be used, in principle, to knock out one by one Majorana fermions bound to
the cores of isolated vortices.

%%%%%% BEGIN FIGURE
%%%%%% END FIGURE

\section{
Conclusions
        }

In summary, we have identified a mechanism to overcome the
fermion-doubling barrier that can prevent the attachment of an odd
number of Majorana fermions to the core of SC vortices in
graphenelike tight-binding models. This mechanism relies on a
$\mathbb{Z}^{\ }_{2}\times\mathbb{Z}^{\ }_{2}$ topological charge that
measures the parity in the number of Majorana fermions attached to an
isolated vortex and the use of TRS-breaking order parameters that
compete with each other and with the SC order parameter to knock out
one by one the Majorana fermions. In this surgical way, an odd number
of Majorana fermions can be made to bind the vortices in a singlet SC
order parameter, whereas this could only be achieved for the more
elusive triplet SC order parameter in Refs.~\onlinecite{Read00},
\onlinecite{Ivanov01}, and \onlinecite{Sato09}. 
This mechanism applies to graphene
with superconductivity induced by the proximity effect, provided a way
can be found to also induce and select from the remarkably large
variety of coexisting and competing order parameters that 
graphene supports those with odd 
$\mathbb{Z}^{\ }_{2}\times\mathbb{Z}^{\ }_{2}$ 
topological charge and thus odd number
of Majorana fermions attached to isolated vortices.

\textbf{ACKNOWLEDGMENTS}

This work is supported in part by the DOE under Grant No. DE-FG02-06ER46316 
(C.C.).
C.M. and S.R. thank the Condensed Matter Theory Visitor's Program
at Boston University for support.


\begin{thebibliography}{99}
\bibitem{footnotethermal}
When $\Delta\neq0$, PHS symmetry is broken
so that the electron charge is not conserved and
it is meaningless to define a Hall conductivity for the
charge when the Haldane masses are dominant. 
Nevertheless, a quantized thermal Hall effect is meaningful
when the Haldane masses are dominant,
for the Bogoliubov-de-Gennes Hamiltonian is conserved,
i.e., the flow of quasiparticle energy is a conserved one.
When $\Delta=0$ and for nonvanishing
Haldane masses, the electron and Nambu quasiparticle charges
are independently conserved. The charge and thermal Hall
conductivities can then be quantized simultaneously.
\end{thebibliography}
\end{document}